\documentclass{JHEP3}
\usepackage{amsmath}
\usepackage{amssymb}

\usepackage{epsfig}

\newcommand{\GeV}{\mathrm{GeV}}

\def\as{\alpha_{{\textsc{s}}}}

\title{Angular ordering and parton showers for non-global QCD
  observables}
       
  \author {Andrea Banfi \\
  Universit\`a degli Studi di Milano-Bicocca\\ and INFN,
  Sezione di Milano-Bicocca, Italy.\\
E-mail: \email{Andrea.Banfi@mib.infn.it}}
  \author{Gennaro Corcella\\
  Dipartimento di Fisica, Universit\`a di Roma `La Sapienza'\\
  Piazzale A. Moro 2, I-00185 Roma, Italy.\\
E-mail: \email{Gennaro.Corcella@roma1.infn.it}}
  \author{ Mrinal Dasgupta\\
  School of Physics and Astronomy, University of Manchester \\
  Oxford  Road, Manchester M13 9PL, U.K.\\
E-mail: \email{Mrinal.Dasgupta@manchester.ac.uk}}

\abstract{ We study the mismatch between a full calculation of
  non-global single-logarithms in the large-$N_c$ limit and an
  approximation based on free azimuthal averaging, and the consequent
  angular-ordered pattern of soft gluon radiation in QCD. We compare
  the results obtained in either case to those obtained from the
  parton showers in the Monte Carlo event generators HERWIG and
  PYTHIA, with the aim of assessing the accuracy of the parton showers
  with regard to such observables where angular ordering is merely an
  approximation even at leading-logarithmic accuracy and which are
  commonly employed for the tuning of event generators to data.  }

\preprint{
Bicocca-FT-06-19\\
MAN/HEP/2006/38\\
ROME1/1444/06\\
hep-ph/0612282}


\begin{document} 


\section{Introduction}
An important class of theoretical predictions in QCD fall under the
banner of ``all-order'' calculations. This refers specifically to
predictions for those observables that receive logarithmic
enhancements at each order of perturbation theory, which threaten the
convergence of the perturbation expansion in important regions of
phase space. A classic example is event-shape distributions, where
studying an observable close to its Born value (such as the
distribution of the thrust variable $1-T$ near $T=1$) results in
generating terms as singular as $\alpha_s^n \frac{1}{1-T}\ln^{2n-1}
(1-T)$ in the perturbative prediction, which can render all orders in
$\alpha_s$ equally significant \cite{CTTW}. The origin of these
logarithmic enhancements is the singular behaviour of the QCD emission
probabilities and their virtual counterparts in the soft and/or
collinear kinematical regions.  These singularities coupled with the
nature of the observable (where measuring close to the Born value
constrains real emission but not the purely virtual terms) lead to the
appearance of large uncancelled logarithmic contributions in the
fixed-order perturbative results.

There exist two main approaches to deal with such logarithmic
enhancements at all orders.  The first is the method of analytical
resummation where insight on the QCD multiple soft-collinear emission
probabilities and analytical manipulations of the phase space
constraints are carried out\footnote{There exist a variety of formal
  approaches designed to achieve these goals all of which embody the
  physics that we outline here.}  so as to obtain a result that resums
the large logarithms (for those variables that satisfy certain
conditions ensuring they can in fact be resummed \cite{BSZ}) into a
function which can be expressed in the form
\begin{equation}
\label{eq:expo}
  \Sigma(V) = 
  \exp \left[Lg_1 (\alpha_s L) + 
    g_2 (\alpha_s L) +
    \alpha_s g_3(\alpha_s L) + \cdots \right] \,,\qquad L \equiv 
\ln \frac{1}{V},
\end{equation}
where $L g_1$, $g_2$, etc. are functions that are computed
analytically\footnote{We include in the category of ``analytical'' the
  semi-analytical approach of Ref.~\cite{numsum} where analytical
  observations are exploited such that $g_2$ can be calculated
  numerically in an automated fashion for several observables.}  and
$V$ is a generic event shape, e.g.  $1-T$.  The function $L g_1$, if
non-zero, represents the leading or double-logarithmic contribution
(LL), since it contains an extra power of $L$ relative to the power of
$\alpha_s$, i.e. ${\cal O}(\alpha_s^nL^{n+1})$.  $g_2$ is the
single-logarithmic or next-to-leading logarithmic (NLL) contribution
containing a logarithm $L$ for each power of $\alpha_s$, ${\cal
  O}(\alpha_s^nL^n)$, etc. We especially note that if the function
$g_1$ is zero (as in the case of the interjet energy flow observable
we shall study in detail here), the single-logarithmic function $g_2$
contains the leading logarithms.  The function $\alpha_s g_3$ contains
an extra power of $\alpha_s$ relative to the power of $L$ and is
next-to--next-to leading logarithmic (NNLL) if $g_1$ is present and
next-to--leading logarithmic (NLL) otherwise. In the limit $V\!\to 0$,
$\Sigma(V)$ has a physical behaviour as opposed to its expansion to
any fixed order, which is divergent as we mentioned. This expression,
which is valid at small $V$, can then be matched to exact fixed-order
estimates that account for the large-$V$ region, so as to give the
best possible description over the entire range of $V$.

Another possible approach to studying such observables is provided by
Monte Carlo event generators amongst which the most commonly employed
are HERWIG~\cite{herwig,herwig++} and PYTHIA~\cite{pyold,pythia,
  pytnew}. We note that these programs are of far greater general
utility than the study of the observables we will discuss here,
providing simulations of complete QCD events at hadron level and
representing perhaps the most significant physics tools in current
high-energy phenomenology.  The parton showers contained in these
event generators aim to capture at least the leading infrared and
collinear singularities involved in the branching of partons, to all
orders in the large-$N_c$ limit. One may thus expect that the dynamics
that is represented by the parton shower ought to be similar to that
which is used as analytical input in QCD resummations at least on the
level of the leading (double) logarithms involved.

For several observables a correspondence between the Monte Carlo
parton shower and the matrix elements used in analytical resummations
is in fact clear.  Considering, for example, final-state radiation,
parton showers evolve due to parton emission with the branching
probability $\mathcal{P}$ satisfying \cite{sudakov,pytorig,mw}
\begin{equation}
\label{eq:herw}
\frac{d {\mathcal{P}}}{d \ln k^2 dz}= \frac{\as}{2\pi}\,   P(z)\, 
\frac{ \Delta(k^2_{\mathrm{max}},k_0^2)}{\Delta(k^2,k_0^2)}\,,
\end{equation}
where $k^2_{\max}$ is the maximum $k^2$ accessible to the
branching and $k^2_0$ is a cut-off regularising soft and collinear
singularities. The above result, with $P(z)$ being the appropriate
Altarelli--Parisi splitting function relevant to the branching,
captures the soft ($z\to 0$) and collinear ($k^2\to 0$)
singularities of the emission. Virtual corrections (and hence
unitarity) are incorporated via the Sudakov form factors
$\Delta(k^2,k_0^2)$.

An essentially similar form is employed for the purposes of most
analytical resummations where the probability of emitting several soft
gluons is treated as independent emission of the gluons by the hard
partons which for simplicity, in the rest of this paper, we take to be
a $q \bar{q}$ pair.  The probability for emitting a soft and/or
collinear gluon is the very form mentioned above and the virtual
corrections are included as in the Sudakov factor.  This
independent-emission or probabilistic pattern (which stems from the
classical nature of soft radiation) suffices up to next-to--leading or
single logarithmic accuracy for a large number of observables.  Thus
it is natural to expect that at least as far as the double-logarithmic
function $g_1$ is concerned, it would be accurately contained within
the parton shower approach, although it cannot be separated cleanly
from the single-logarithmic and subleading effects generated by the
shower.  Beyond the double logarithmic level one expects at least a
partial overlap between the parton shower and the analytical
resummations, where the degree of overlap may vary from observable to
observable and depend on which hard process one chooses to address.
The state of the art of most analytical resummations is
next-to--leading logarithmic, i.e. computing the full answer up to the
function $g_2$. Monte Carlo algorithms such as HERWIG are certainly
correct up to $g_1$ and perhaps in certain cases $g_2$ accuracy (while
being limited to the large-$N_c$ approximation) but not beyond (see,
e.g., the discussion in \cite{cmw}).

As we mentioned, the event generator results do not explicitly
separate leading logarithmic from next-to-leading logarithmic or
subleading effects (e.g. those that give rise to $g_3$ and beyond)
and, moreover, parton-level Monte Carlo results include
non-perturbative effects that arise via the use of a shower cut-off
scale, i.e. $k_0$ in Eq.~(\ref{eq:herw}).  From the point of view of
having a clean prediction valid to NLL accuracy that can be matched to
fixed-order and supplemented by, for instance, analytically estimated
power corrections, one would clearly prefer a resummed calculation.
This is not a surprise since these calculations were developed keeping
specific observables in mind unlike the event generators which have a
much broader sweep and aim. It is thus not our aim to probe event
generators as resummation tools in themselves but rather to consider
the logarithmic accuracy to which perturbative radiation may be
generically described by a parton shower of the kind to be found in
HERWIG or PYTHIA, for different observables.

The above is particularly important since it has been pointed out
relatively recently that for a large number of commonly studied
observables, which are called non-global observables
\cite{DassalNG1,DassalNG2}, the approximation of independent
emissions, used in the analytical resummations, is not valid 
to single (which for some of these observables means leading)
logarithmic accuracy.  Non-global observables typically involve
measurements of soft emissions over a limited part of phase space, a
good example being energy flow distributions in a fixed
rapidity-azimuth ($\eta-\phi)$ region.  In fact in the case of the
energy flow away from hard jets the function $g_1$ in
Eq.~\eqref{eq:expo} is absent (there being no collinear enhancement in
the away-from-jet region). The leading logarithms in this case are 
thus single logarithms that are resummed in a function equivalent to $g_2$
but this function cannot be completely calculated within an
independent emission formalism. This is the case because the
independent emission approximation of the QCD multi-parton emission
pattern is strictly valid and intended for use in regions where
successive emissions are strongly ordered in angle. The leading
partonic configurations (those that give rise to the leading
single-logarithms) for the away-from--jet energy flow are however
those which include the region of emission angles of the same order in
the parton cascade.  Thus relevant single-logarithms also arise from
multi-soft correlated emission which has been computed only
numerically and in the large-$N_c$ limit thus
far~\cite{DassalNG1,DassalNG2,BMS}.

Since one of the main approximations used in analytical resummations,
that of independent emission, has been shown to be inaccurate even to
leading-logarithmic accuracy for some non-global observables like
interjet energy flow, one is led to wonder about the
leading-logarithmic accuracy that is claimed for parton showers in
Monte Carlo event generators, in these instances. The parton shower in
HERWIG for instance relies on an evolution variable $k^2$ which in the
soft limit is equivalent to ordering in angle~\cite{mw,mw88}. Angular
ordering of a soft partonic cascade, initiated by a hard leg, is a
perfectly good approximation for azimuthally averaged quantities such
as some $e^{+}e^{-}$ event shapes and in fact can be further reduced
in these instances to an independent emission pattern, up to
next-to--leading logarithmic accuracy.  However, when looking at
energy flow into limited angular intervals, one is no longer free to
average soft emissions over the full range of angles, which means that
one no longer obtains angular ordering at single-logarithmic accuracy.
Thus one expects at least formally that the parton shower in HERWIG is
not sufficient even to leading logarithmic accuracy for variables such
as energy flow in inter-jet regions. The same statement should apply
to the PYTHIA shower and even more strongly to versions before 6.3
where the ordering variable is always taken as the virtuality or
invariant mass and angular ordering imposed thereafter \cite{pyold}, 
which leads to
insufficient phase-space for soft emission. Version 6.3
\cite{pythia} offers as an alternative the possibility to order the
shower according to the transverse momentum of the radiated parton
with respect to the emitter's direction (see \cite{pytnew} for more discussion
on the transverse momentum definition), which yields a better
implementation of angular ordering \cite{pytnew}.  We would like to
point out that the ARIADNE Monte Carlo generator \cite{ariadne} has
the correct large-angle soft gluon evolution pattern, which generates
the non-global single logarithms in the large-$N_c$ limit.  Since
however the most commonly used and popular programs are the ones we
mentioned before, we shall be interested in comparisons to the showers
therein.

This issue assumes some importance while considering for instance the
tuning of the shower and non-perturbative parameters in Monte Carlo
generators.
If the tuning is performed by using data on a non-global observable
such as energy flow away from jets one must at least be aware of what
the accuracy is of the shower produced by the event generator. If the
accuracy is not even leading-logarithmic then one runs the risk of
incorporating missing leading-logarithmic effects via tuned
parameters. This situation is not optimal since, as far as possible, one
would like to account only for subleading effects and incalculable
non-perturbative physics via the tuning.  Moreover, the soft physics of
non-global observables is not universal, the multi-soft correlated
emission component being irrelevant in the case of global observables
(those sensitive to soft emission over the full angular range).  This
difference in sensitivity to soft gluons, for different observables,
would not be accounted for in case the non-global effects are tuned in
once and for all.

In the present paper we aim to investigate the numerical extent of the
problem and to what extent non-global logarithms may be simulated by
angular ordering and hence by parton shower Monte Carlo generators.
In the following section we shall compare a fixed order
${\mathcal{O}}(\alpha_s^2)$ calculation of the leading non-global
effect for energy flow into a rapidity slice with that from 
a model of the
matrix element where we impose angular ordering. We shall comment on
the results obtained and in the following section examine what happens
at all orders and whether our fixed-order observations can be
extrapolated. Having compared the full non-global logarithmic
resummation with its angular-ordered counterpart we then proceed to
examine if our conclusions are borne out in actual Monte Carlo
simulations. Thus we compare the results of resummation with those
obtained from HERWIG and PYTHIA at parton level. This helps us arrive
at our conclusions on the role of non-global effects while comparing
Monte Carlo predictions to data on observables such as the energy flow
between jets, which we report in the final section.

\section{Non-global logarithms vs angular ordering at leading order}

In order to explore the issues we have raised in the introduction, we
pick the interjet energy flow (more precisely transverse energy $E_t$
flow) observable.  Here there are no collinear singularities and the
problem reduces to one where the leading logarithms encountered in the
perturbative prediction are single-logarithms.  While the nature of
the hard-process is fairly immaterial in the large-$N_c$ limit to
which we confine our discussions, it proves simplest to choose
$e^{+}e^{-}\!\to\! 2$ jets and examine the $E_t$ flow in a chosen angular
region.

Given a phase-space region $\Omega$, the $E_t$ flow is defined as
\begin{equation}
  \label{eq:Et}
  E_t = \sum_{i\in\Omega} E_{ti}\,,
\end{equation}
where the sum runs over all hadrons (partons for our calculational purposes) 
and the observable we wish to study is
\begin{equation}
  \label{eq:sigma-def}
  \Sigma(Q,Q_\Omega) = \frac{1}{\sigma}\int_0^{Q_\Omega} \!\!\!dE_t 
  \frac{d\sigma}{d E_t}\,.
\end{equation}

The theoretical result for the integrated quantity $\Sigma$ was correctly 
computed to single-logarithmic accuracy in Ref.~\cite{DassalNG2} and assumes 
the form 
\begin{equation}
\label{eq:Sigma}
 \Sigma(Q,Q_\Omega) = \exp[-4 C_F A_\Omega t]\> S(t),
\end{equation}
where one has defined $t$
\begin{equation}
  \label{eq:tdef}
  t(L)=\int_{Qe^{-L}}^{Q} \frac{dk_t}{k_t}\frac{\as(k_t)}{2\pi} \,,
  \qquad L \equiv \ln \frac{Q}{Q_\Omega}.
\end{equation}
The first factor in Eq.~\eqref{eq:Sigma} above is essentially a
Sudakov type term where $A_\Omega = \int d\eta \frac{d\phi}{2 \pi}$
represents the area of the region $A_\Omega$. Note the colour factor
$C_F$ from which it should be clear that this term is related to
multiple independent emission off the hard primary $q \bar{q}$ pair
and in fact is just the exponential of the single-gluon emission
result.

The second factor $S(t)$ is the correlated gluon emission contribution
which starts with a term that goes as $C_F C_A \alpha_s^2 \ln^2
(Q/Q_\Omega)$.  This can be calculated fully analytically while the full
resummed single-logarithmic calculation for $S(t)$ is carried out
numerically in the large $N_c$ limit. Before we turn to the all-orders
result we aim to compare the analytical leading-order computation with
a model of the matrix element based on angular ordering. This will
give us some insight into the issue at hand.

In order to do so we start with the full matrix-element squared for
energy-ordered two gluon emission from a $q \bar{q}$ dipole $ab$:
\begin{multline}
\label{eq:me}
M^2(k_1,k_2) = 4 C_F\> \frac{(ab)}{(ak_1)(bk_1)} \>\times \\
\times
\left [ \frac{C_A}{2} 
  \frac{(ak_1)}{(ak_2)(k_1k_2)}+\frac{C_A}{2} \frac{(bk_1)}{(b k_2)(k_1k_2)}+
  \left(C_F-\frac{C_A}{2} \right) 
  \frac{(ab)}{(ak_2)(bk_2)} \right]\,,
\end{multline}
with the conventional notation $(a b) = a\!\cdot\! b$ with $a$, $b$ and
$k_i$ being the particle four-momenta.  We define these four-vectors
as below:
\begin{eqnarray}
\label{eq:mom}
a & =& \frac{Q}{2} \left (1,0,0,1 \right )\,, \\ \nonumber
b &= &\frac{Q}{2} \left (1,0,0,-1 \right )\,,\\ \nonumber
k_1 &=& k_{t,1} 
\left (\cosh \eta_1, \cos \phi_1, \sin \phi_1, \sinh \eta_1 \right )\,,
\\ \nonumber
k_2 &= & k_{t,2} 
\left (\cosh \eta_2, \cos \phi_2, \sin \phi_2, \sinh \eta_2 \right)\,,
\end{eqnarray}
where $Q$ is the centre-of--mass energy. 

We also separate the ``independent emission'' piece of the squared
matrix element, proportional to $C_F^2$, from the correlated emission
piece proportional to $C_F C_A$:
\begin{equation}
M^2(k_1,k_2) =  C_F^2 \>W(k_1) W(k_2) +C_F C_A \>W(k_1,k_2)\,.
\end{equation}
It is this latter piece that is
termed the non-global contribution at this order.

We now wish to distinguish between a full calculation of the
non-global contribution at $\mathcal{O} \left(\alpha_s^2 \right)$ and
that based on an angular-ordered model of the squared matrix element.
We first revisit the full result without angular ordering.
Since only the $C_F C_A$ piece of the result will be different in the
angular-ordered approximation, we shall focus on this term. 
Using the
momenta defined in Eq.~\eqref{eq:mom} we obtain
\begin{equation}
  C_F C_A W(k_1,k_2) = \frac{128 C_F C_A}{Q^4 x_1^2 x_2^2} 
  \left [\frac{\cosh \left (\eta_1-\eta_2 \right)}
    {\cosh \left (\eta_1-\eta_2\right) -\cos (\phi_1-\phi_2)}-1\right]\,,
\end{equation}
where we introduced the transverse-momentum fractions $x_i = 2
k_{t,i}/Q$, and assume that $x_1 \gg x_2$, i.e. strong ordering of the 
transverse momenta.

The non-global contribution is given by integrating the above result 
over the
directions of the two gluons such that the softer gluon ($k_2$) 
is in $\Omega$ while the harder gluon ($k_1$) is outside, and over the 
scaled transverse momenta $x_1$ and
$x_2$.  The integral over directions (including a phase space factor $Q^4/16$) 
is given by
\begin{equation}
  C_F\, C_A\, \frac{Q^4}{16}
  \int_{k_1 \notin \Omega}
  \!\!d \eta_1 \frac{d\phi_1}{2\pi}
  \int_{k_2 \in \Omega} 
  \!\!d \eta_2 \frac{d\phi_2}{2\pi}\>\>
  W(k_1,k_2)\,.
\end{equation}
Integrating over the energy fractions $x_1$ and $x_2$ 
produces at the leading single-logarithmic level
a factor $-(1/2) \ln^2 (Q/Q_\Omega)$. The coefficient of
the $\left(\frac{\alpha_s}{2 \pi}\right)^2 \ln^2 (Q/Q_\Omega)$ term has
a $C_F C_A$ or non-global contribution which reads
\begin{equation}
  S_2 =   -4\, C_F\, C_A 
  \int_{k_1 \notin \Omega}
  \!\!d \eta_1 \frac{d\phi_1}{2\pi}
  \int_{k_2 \in \Omega} 
  \!\!d \eta_2 \frac{d\phi_2}{2\pi}\>\>
  \left [\frac{\cosh \left (\eta_1-\eta_2 \right)}
    {\cosh \left (\eta_1-\eta_2\right) -\cos (\phi_1-\phi_2)}-1\right].
\end{equation}

We now choose $\Omega$ as a slice in rapidity of width
$\Delta \eta$ which one can centre on $\eta=0$ with its edges at
rapidities $-\Delta\eta/2$ and $\Delta\eta/2$. We are free to take
$\phi_1 = 0$ and integrating over $\phi_2$ gives the result
\begin{equation}
  S_2 = -8 \,C_F\, C_A \int_{- \infty}^{-\Delta \eta/2} 
  \!\!\!\!d\eta_1
  \int_{-\Delta \eta/2}^{\Delta \eta/2} 
  \!\!d \eta_2
  \,\left [\coth \left (\eta_2-\eta_1 \right)-1 \right]\,,
\end{equation}
where we doubled the result of assuming $\eta_1 < \eta_2$ to account for 
the region $\eta_1 > \eta_2$.

Now one is left with the integral over the gluon rapidities. In order
to examine the main features of the final result, which were already
elaborated in Ref.~\cite{DassalNG1}, we introduce the rapidity
difference $y = \eta_2 -\eta_1$ in terms of which one can reduce the
above integral to 
\begin{equation}
  S_2 =-8\, C_F\, C_A \left (\int_0^{\Delta \eta} \!\!dy\, y \left( \coth y -1 \right ) + 
    \int_{\Delta \eta}^{\infty} \!\!dy\, \Delta \eta \left (\coth y-1 \right) \right).
\end{equation}
Let us concentrate on the case of a large slice where the result has
an interesting behaviour. As one increases $\Delta \eta$ the second
integral in the sum above, from $\Delta \eta$ to infinity, starts to
become progressively less significant. This is because the integrand
$\coth y-1$ rapidly approaches zero as $y$ becomes large.  The first
term in the parentheses, on the other hand, gets its main contribution
from the small $y$ region.  Its value as $\Delta \eta \to \infty$
tends to $\pi^2/12$.  Thus what one observes as one increases $\Delta
\eta$ is that the contribution to the integral from $\Delta \eta \to
\infty$ starts to be negligible while the contribution of the integral
from zero to $\Delta \eta$ starts to become insensitive to its upper
limit and hence the slice width $\Delta \eta$, being dominated by the
contribution from the small $y$ region. This leads to a rapid
saturation of the result as one increases $\Delta \eta$ and the result
quickly approaches $\pi^2/12$.  For instance the value at $\Delta\eta =2.5$
is $0.818$ while $\pi^2/12 = 0.822$.

Now we recompute the above integral using an angular-ordered
approximation of the squared matrix element.  We expect that the
angular ordering we introduce here should correspond to the
contribution to the non-global logarithms that ought to be contained
in Monte Carlo event generators based on angular ordering.  The
angular-ordered approximation to the matrix element squared Eq.~\eqref{eq:me}
is obtained by modifying each dipole emission term therein as below:
\begin{multline}
\label{eq:mod}
  \frac{(ab)}{(ak)(bk)}= \frac{1-\cos\theta_{ab}}{\omega^2
    \left(1-\cos\theta_{ak}\right)\left(1-\cos\theta_{kb}\right)}  \\
  \to \frac{1}{\omega^2} \left (\frac{\Theta\left(\cos \theta_{ak}-
        \cos \theta_{ab} \right)}{1-\cos\theta_{ak}} +\frac{\Theta
      \left (\cos \theta_{kb}-\cos \theta_{ab}
      \right)}{1-\cos\theta_{kb}} \right)\,,
\end{multline}
where $\omega$ refers to the energy of $k$.  The second line above is
actually equivalent to the full result if one can integrate freely
over the azimuthal angles defined with respect to each of the legs of
the emitting dipole, leaving a dependence on just the polar angles
$\theta$.  However, since one places geometrical restrictions on the
emissions $k_1$ and $k_2$, and in that respect $k_1$ has to be outside the gap
while $k_2$ inside, the azimuthal integration does not extend from
zero to $2\pi$.  The limits instead depend on the precise gap
geometry. Ignoring this we wish to model the full matrix element
squared by the angular pattern introduced above, corresponding to
emission of soft gluons in well-defined cones around each hard
emitting leg.

\FIGURE{
 \epsfig{file=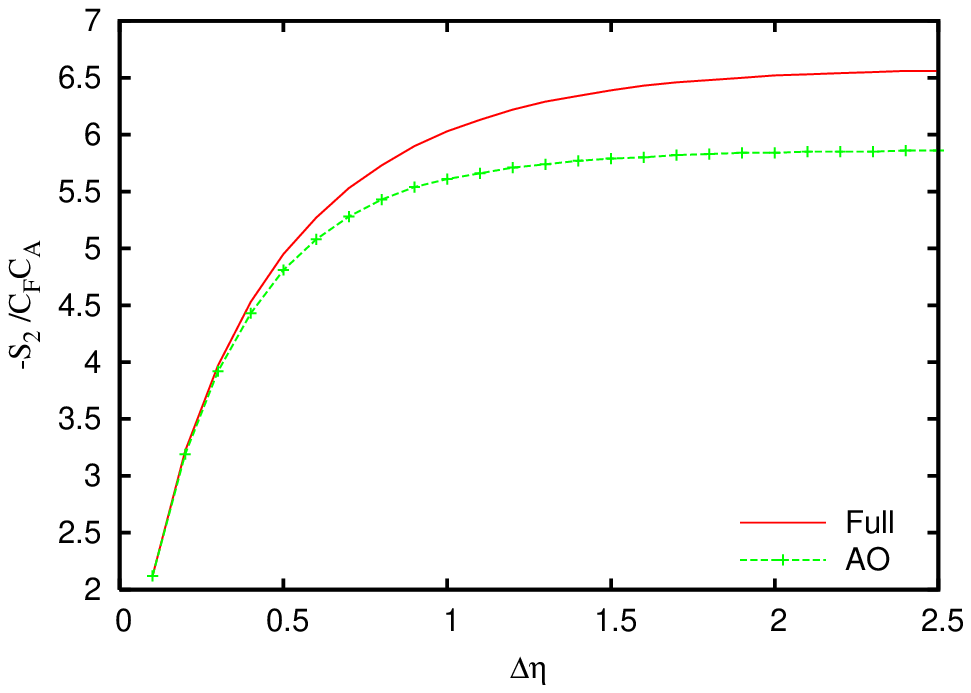, width=.7\textwidth}
 \caption{The coefficient of the leading order non-global contribution
   $-S_2/C_FC_A$ plotted as a function of the rapidity slice $\Delta
   \eta$ as given by both the full calculation and the angular-ordered
   approximation. The significant feature of saturation of the result
   for large slice-widths is apparent in both results.}
\label{fig:ss2}
}

We note once more that 
the $C_F^2$ independent-emission term of the squared matrix
element is left intact since the angular-ordered and full results are
identical for this piece, as one would expect. 
Making the modification described in Eq.~\eqref{eq:mod} in each
term of the $C_F C_A$ piece of the 
squared matrix element Eq.~\eqref{eq:me} and integrating over
gluon directions we obtain the coefficient $S_2$ in the ``angular
ordered'' (AO) approximation.
We plot the numerical result in this approximation 
as a function of the gap size in
Fig.~\ref{fig:ss2} along with the full result. 
One can immediately observe that for small gap
sizes the AO and full results are essentially identical. As one
increases the gap size one notes a numerically significant difference
between the two results although this is at best moderate.  For
instance for a slice of width $\Delta \eta = 2.5$ one observes that
the AO result is lower by $10.67 \%$ than the full result.
Additionally it is interesting to observe that the notable feature of
saturation of $S_2$ for a large gap size is preserved by the AO
approximation.

 The reason the saturation property is preserved is because, as
 explained previously in detail, it arises from the region where the
 two gluons (respectively in and outside the gap region) are close in
 angle or equivalently from the region of integration $\eta_1\! -\!\eta_2
 \leq \Delta \eta$. Moreover, the bulk of the non-global contribution
 for any gap size arises from the region where the emission angles of
 the two soft gluons are of the same order. The contribution from
 configurations with the softest gluon at large angle relative to the
 next-softest gluon are small and vanish rapidly as we make the rapidity
 separation $\eta_1\!-\!\eta_2$ large.

 In the AO approximation one requires the softest gluon $k_2$ to be
 emitted in a cone around the hard emitters $k_1$ and either the
 emitting quark leg $a$ or $b$, depending on whether one is looking at
 emission by dipole $ak_1$ or $bk_1$. The size of the cone is equal to
 the dipole opening angle. Thus the important region where $k_1$ and
 $k_2$ are collinear is perfectly described by the AO model. Only the
 region where $k_2$ is emitted at an angle larger than the cone
 opening angle would not be covered in the AO approximation and the
 contribution of such a region should be relatively small as we
 observe numerically. We mention in passing that these conclusions
 described explicitly for a rapidity slice are expected to hold for a
 general gap geometry and we explicitly checked the case of a square
 patch $\Delta \eta =\Delta \phi$ in rapidity and azimuth.

In the following section we shall examine the impact of the AO
 approximation at all orders to determine whether the encouraging
 fixed-order finding, that an AO model reproduces the characteristics
 and is numerically reasonably close to the full non-global result,
 can be extended to all orders, as one may now expect.

\section{AO approximation at all orders}
We now study the AO approximation by using the large $N_c$ 
evolution algorithm
that was described in Ref.~\cite{DassalNG1}, suitably modifying it to
take account of the angular ordering requirement. This should enable
us to estimate how non-global logarithms will be simulated in an
angular-ordered parton shower event generator.  The algorithm works as
follows. To compute the non-global contribution $S(\alpha_s L)$ where
$L \equiv \ln (Q/Q_\Omega)$ one considers the probability $P_C(L)$ of a
configuration $C$ that does not resolve gluons above scale $L$, in
other words those with energies below $Qe^{-L}$.  The evolution of
this configuration to another configuration $C'$ with larger
resolution scale $L'$ or equivalently smaller energy scale, proceeds
via soft emission of an extra gluon $k'$ from the configuration $C$:
\begin{equation}
\label{eq:evol}
P_{C'} (L') = \bar{\alpha}_s(L') \Delta_C(L,L')P_C (L) F_C (\theta',\phi')\,,
\end{equation}
where $\Delta_C(L,L')$ represents the summation of only virtual gluons
between the scales $L$ and $L'$, $F_C (\theta',\phi')$ represents the
angular pattern of emission of gluon $k'$ from the system of dipoles
in the configuration $C$ and $\bar{\alpha}_s \equiv
\alpha_s/(2\pi)$. One has explicitly
\begin{equation}
  F_C (\theta_k,\phi_k) = \sum_{\mathrm{dipoles-ij}}
  \frac{2 C_A \left(1-\cos\theta_{ij}\right)}{\left(1-\cos\theta_{ik}\right)
    \left(1-\cos\theta_{jk}\right)}.
\end{equation}
The same dipole angular pattern enters the pure virtual evolution
probability (or form factor):
\begin{equation}
\ln \Delta_{C}(L,L')= - \int_L^{L'}\!\! d L'' \!\int d \cos \theta \>d \phi 
\>\bar{\alpha}_s (L'')\> F_C (\theta, \phi).
\end{equation}

The probability that the interjet region $\Omega$ stays free of real
emissions below a given scale $L$, is then given by summing over
corresponding dipole configurations:
\begin{equation}
  \Sigma(Q,Q_\Omega) = \sum_{C \,|\, \Omega\, \mathrm{empty}} P_C(L)\,.
\end{equation}

In order to obtain our angular-ordered results we need to modify the
angular emission pattern $F_C$, as before for the fixed-order case, 
so we define
\begin{equation}
F_C(\theta_k,\phi_k)_{\mathrm{AO}} \equiv \sum_{\mathrm{dipoles-ij}} 
2 C_A \left ( \frac{\Theta
    \left(\cos \theta_{ik}-\cos\theta_{ij} \right)}
  {1-\cos\theta_{ik}} + 
  \frac{\Theta\left(\cos \theta_{jk}-\cos\theta_{ij} \right)}
  {1-\cos\theta_{jk}} \right)\,.
\end{equation}

Making the replacement $F_C( \theta_k, \phi_k) \to
F_{C}(\theta_k,\phi_k)_\mathrm{AO}$ one modifies both real and virtual
terms and obtains the result from our angular-ordered model at all
orders:
\begin{equation}
\label{eq:sigma-ao}
\Sigma_\mathrm{AO}(Q,Q_\Omega) = \sum_{C \,|\, \Omega\, \mathrm{empty}} 
P_{C,\mathrm{AO}}(L)\,.
\end{equation}

Having obtained $\Sigma_\mathrm{AO}$ we can compare it with the full
result without angular ordering.  In Fig.~\ref{fig:Sigmafullvsao} we plot
the full and AO results for $\Sigma(t)$ as a function of $t$, for a
slice of unit width $\Delta \eta =1$.

\FIGURE{
    \epsfig{file=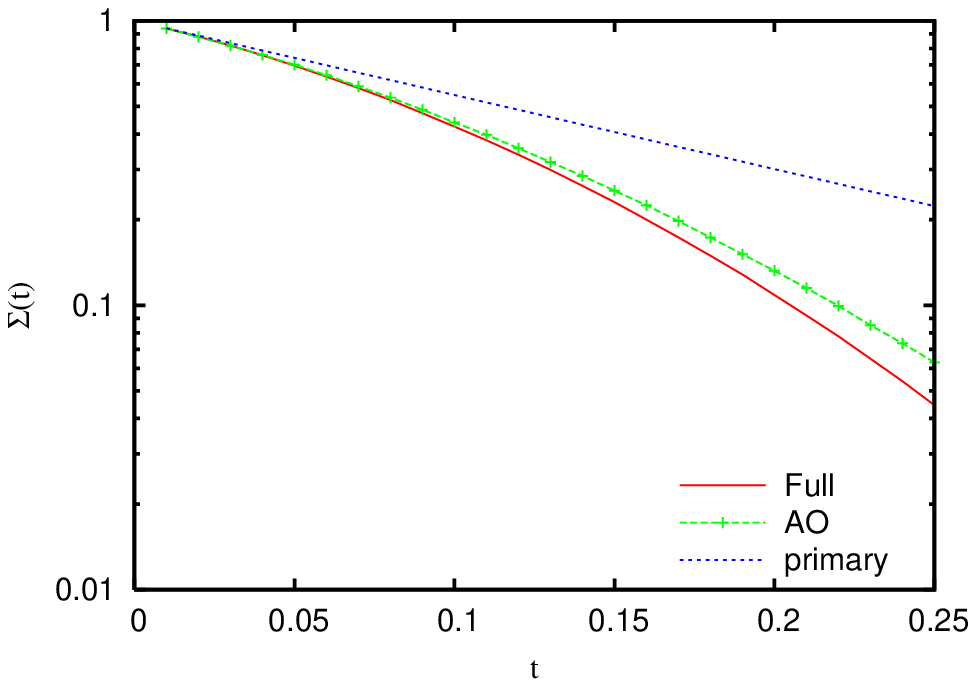, width=.8\textwidth}
    \caption{The integrated cross-section as a function of $t$ in the full
      calculation and the AO approximation.  
      The primary result is also shown for reference.}
    \label{fig:Sigmafullvsao}
}
One notes the relatively minor difference between the full and the AO
results which indicates that the contribution to the full answer from
regions where one can employ angular ordering, is the dominant
contribution.  For the sake of illustration we focus on the value
$t=0.15$ which corresponds to a soft scale $Q_\Omega = 1.0$ GeV for a
hard scale $Q =100$ GeV.  For the rapidity slice of unit width we note
that the result for $\Sigma_{\mathrm{AO}}(t)$ is 9.68 $\%$ higher than
the full result.  At the same value of $t$,
the difference between the full
and the primary result, i.e. $\exp(-4\, C_F \,t)$, is
around 75 $\%$, thus indicating that the AO approximation is much less
significant than the role of the non-global component itself. Similar
observations hold regardless of slice width.

One can also directly study the impact of the AO approximation on the
pure non-global contribution $S(t)$.  The primary contribution is
unaffected by angular-ordering and can be divided out from the result
for $\Sigma_{\mathrm{AO}}(t)$ to give us $S_{\mathrm{AO}}(t)$.  We
first take the example where $\Omega$ is a rapidity slice and consider
different values for the slice width $\Delta \eta$.  We illustrate in
Fig.~\ref{fig:SfullvsAO} three choices for the slice width $\Delta \eta
=1.0, \,2.0, \, 3.0$ with the full non-global contribution $S(t)$ and
that in the AO model. We note that in both full and AO cases the
feature of rough independence on the slice width $\Delta \eta$ is
seen, as one can expect for sufficiently large slices. The AO curves
are somewhat higher than the full ones indicating a somewhat smaller
suppression than that yielded by the full calculation.

\FIGURE{
    \epsfig{file = 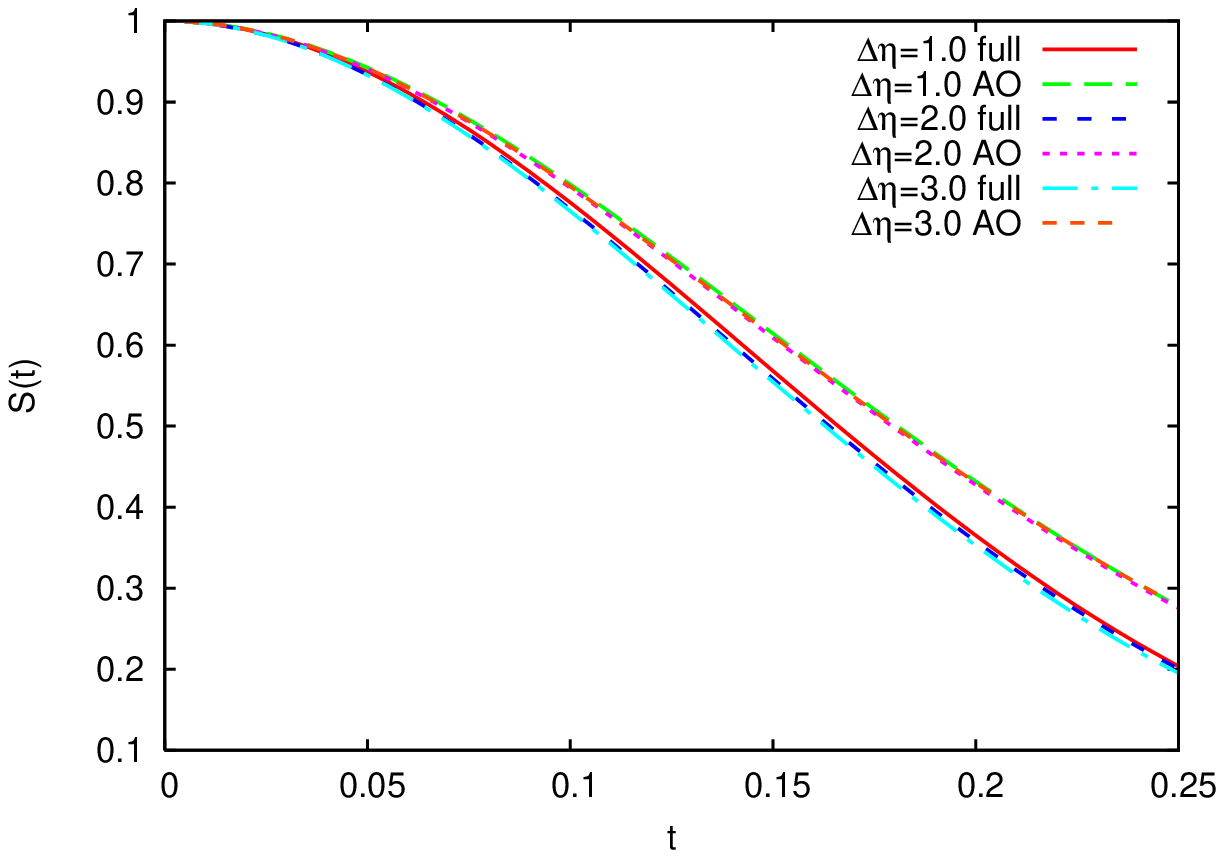, width=.8\textwidth}
    \caption{The resummed non-global contribution $S(t)$ as a function of
      $t$ in the full calculation and the AO approximation for different
      values of the slice width $\Delta \eta$. The upper set of curves
      correspond to the AO case and reflect that in that approximation a
      slightly smaller suppression is obtained than from the full
      calculation corresponding to the lower set of curves. The feature of
      rough independence on the slice width $\Delta \eta$ is visible in
      the full case and is preserved by the AO approximation.}
    \label{fig:SfullvsAO}
}

\FIGURE{
    \epsfig{file=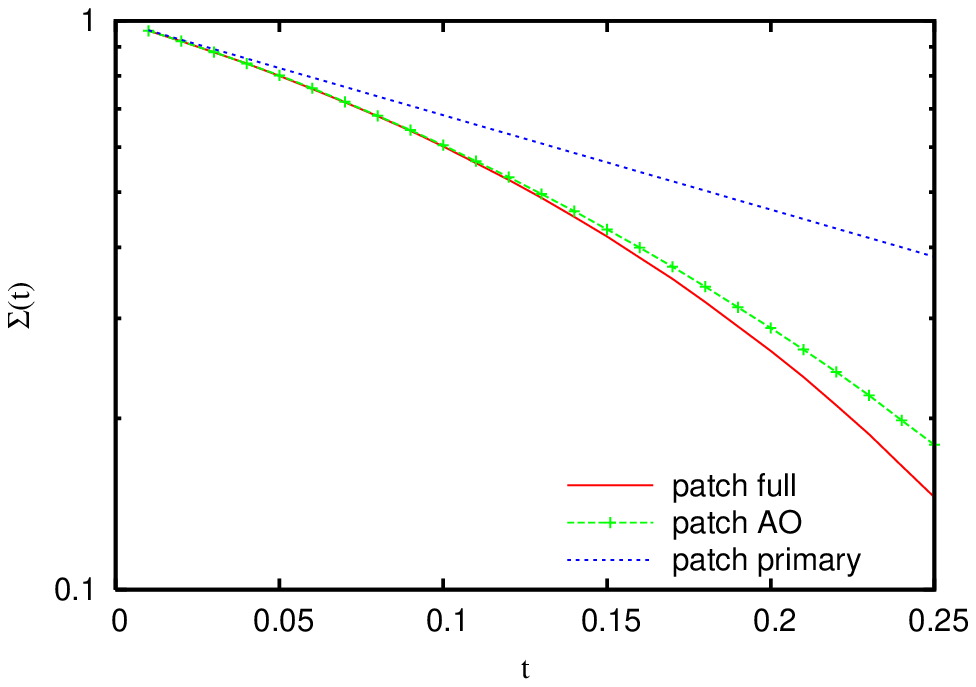, width=.8\textwidth}
    \caption{$\Sigma(t)$ vs $t$ for a square patch in rapidity and
      azimuth, $\Delta \eta = \Delta \phi =2.0$.  Primary, full and
      angular-ordered (AO) curves are shown.}
    \label{fig:patch1}
}

\FIGURE{
    \epsfig{file=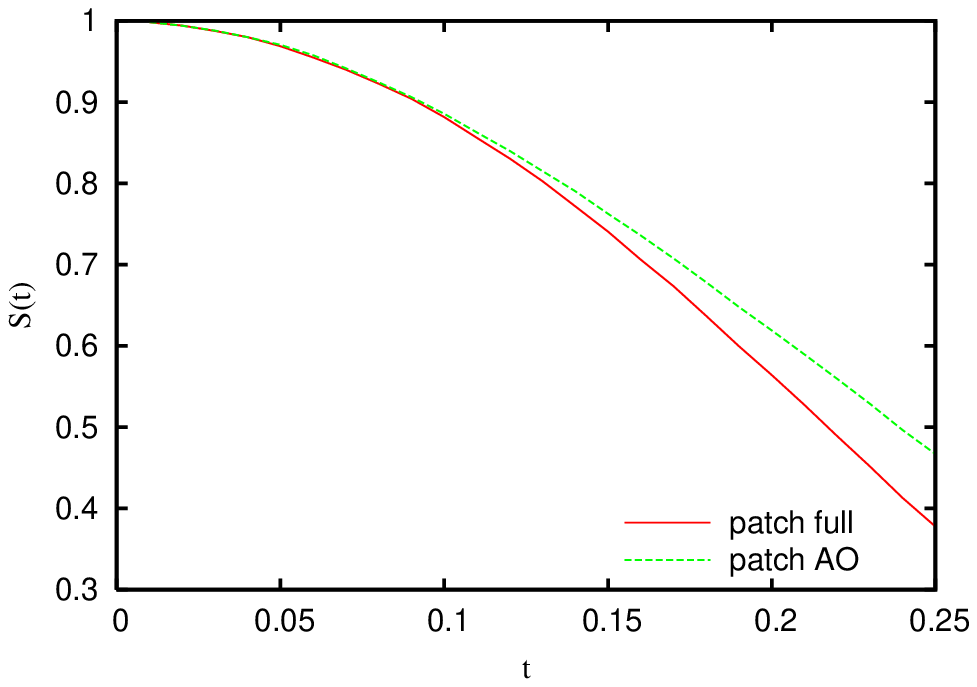, width=.8\textwidth}
    \caption{The non-global contribution $S(t)$ as a function of $t$
      for a square patch in rapidity and azimuth, $\Delta \eta =\Delta
      \phi =2.0$. }
    \label{fig:Sfullvsaopatch}
}

Similar studies can be carried out for different geometries of
$\Omega$.  For a square patch in rapidity and azimuth with $\Delta
\eta = \Delta \phi =2.0 $, the full and angular-ordered results for
$\Sigma(t)$ are shown in Fig.~\ref{fig:patch1}.  The difference is seen to
be small over a wide range of $t$. Once again focusing on the
$t=0.15$ value, one notes that the AO approximation is only about
three percent above the full result.  At $t = 0.2$ this difference
rises to $9.75$ $\%$.  Corresponding results for $S(t)$ for the same
square patch, obtained by dividing by the primary result, are plotted
in Fig.~\ref{fig:Sfullvsaopatch} and once again only a small to moderate
effect is observed over the $t$ range shown.

We have thus observed that modifying the evolution code
\cite{DassalNG1} used to compute the non-global logarithms, to impose
angular ordering on them, only has a moderate effect on the quantity
$S(t)$.  This effect becomes even less significant for the quantity
$\Sigma(t) = \Sigma_P(t)\, S(t)$ since the primary contribution
$\Sigma_P(t)\equiv\exp[-4 C_F A_\Omega t]$ is unchanged by imposing angular
ordering, which we also explicitly checked with the code.

Having thus noted the small effect of the AO approximation within our
model we would not expect much difference, in principle, between the
results from an event generator based on angular-ordering in the soft
limit (HERWIG) and the full non-global results. For PYTHIA, prior to
the version 6.3 one may expect to see differences since angular
ordering was imposed on top of ordering in the virtuality (invariant
mass) of a splitting parton which leads to known problems with
soft-gluon distributions, as discussed in \cite{pytnew}.
In Ref.~\cite{cdf}, where colour coherence effects were observed and
studied at the Tevatron collider, it was in fact found that, unlike
HERWIG, the PYTHIA event generator was not able to acceptably reproduce 
experimental observables sensitive to angular ordering.
One may expect, however, that the new PYTHIA 
model \cite{pythia,pytnew}
(where, to our understanding, the improved shower, ordered in
transverse momentum, better accounts for angular-ordering) 
results comparable to those from HERWIG may be obtained.  
In the next section our aim is to explore these issues and see if our 
expectations, outlined above, are indeed borne out.

\section{Comparison with HERWIG and PYTHIA}
\label{sec:herwig}
In this section we shall focus on actual comparisons to results from
HERWIG and PYTHIA. In order to meaningfully compare the results of a
leading-log resummation with the parton level MC results, it is
necessary to minimise the impact on the MC results of formally
subleading and non-perturbative effects that are beyond full control
and hence spurious.

In order to suppress subleading effects one needs to carry out the
comparisons to the MC generators at extremely high values of $Q$, and hence we
chose $10^5$ GeV.  Thus effects that are formally of relative order
$\alpha_s(Q)$ or higher can be expected to be negligible.  A sign of
this is the fact that at such large $Q$ values the MC results one
obtains do not depend on $Q$ other than via the single logarithmic
variable $t$
for a large range of $t$.  It is clear that such high $Q$ values are
beyond the reach of current or imminent collider experiments but since
we are interested only in the dependence on $t$, the $Q$ value is
fairly immaterial for our purposes.  In fact one can take the
conclusions we make for a particular $t$ value at $Q =10^5$ GeV and
translate that into a value of $Q_\Omega$ for an experimentally
realistic value of $Q$.  

A clear source of uncertainty in this procedure is the
different definitions of $\as$ in the resummation and the MC programs.
In all resummed predictions we have used the LL expression for $t$:
\begin{equation}
  \label{eq:t-LL}
  t = \frac{1}{4\pi\beta_0} \ln\frac{1}{1-2 \as(Q) \beta_0 L}\,,\qquad
  \beta_0=\frac{11 C_A-4 T_R n_f}{12\pi}\>, 
\end{equation}
with $\beta_0$ corresponding to $n_f=6$ and $L$ given in Eq. (\ref{eq:tdef}).
The coupling $\as(Q)$ is in
the $\overline{\mathrm{MS}}$ scheme, and is obtained via a two-loop
evolution with 6 active flavours from the input value $\as(M_Z)=0.118$.
This is to ensure that the resummed prediction is a function of
$\as(Q) L$ only. HERWIG instead exploits a two-loop coupling in the
physical CMW scheme~\cite{cmw} with $\as(M_Z)=0.116$, while PYTHIA
uses a one-loop coupling corresponding to
$\as(M_Z)=0.127$~\cite{pythia}. The values of $t$ corresponding to
different definitions of $\as$ (computed according to
Eq.~(\ref{eq:tdef})) are found to be compatible within 10\% in the
considered $E_t$ range. This does not lead to appreciable
modifications in the resummed curves plotted as a function of $E_t$ 
rather than $t$, in this section. Thus the comparisons we make below 
to the Monte Carlo results at a particular value of $E_t$ are not significantly affected by the 
issue of the somewhat different definitions employed in the resummation and 
the various Monte Carlo programs.

Another effect, not accounted for in the resummation, that is
potentially significant, is the effect of quark masses (which would
arise due to excitation of all flavours). These effects however 
can be safely
neglected at the value of $Q$ we choose. In particular we also 
note that the
presented MC curves are obtained by allowing the top quark to decay,
but we have explicitly checked that we obtain almost identical results
if we force the top quark to be stable.

\FIGURE{
    \epsfig{file=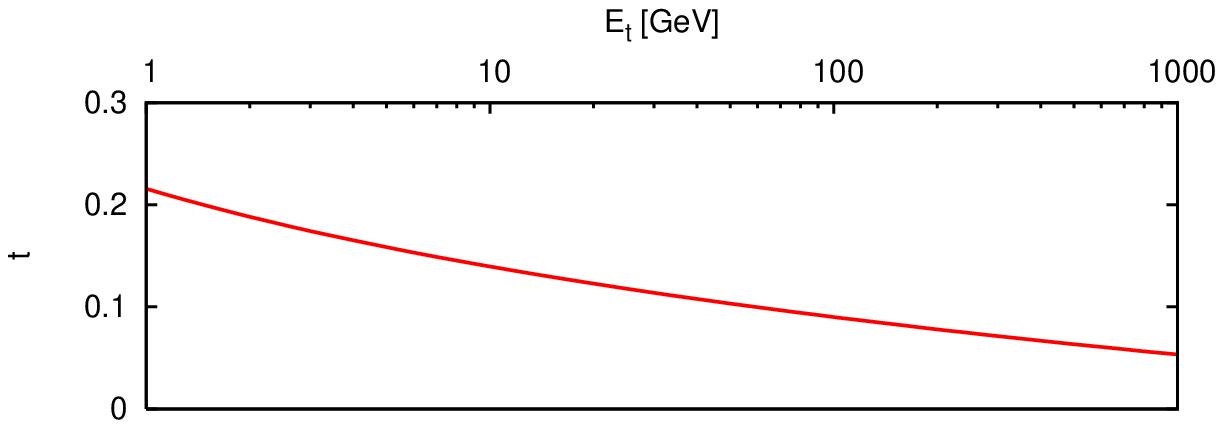,width=\textwidth}\\
    \epsfig{file=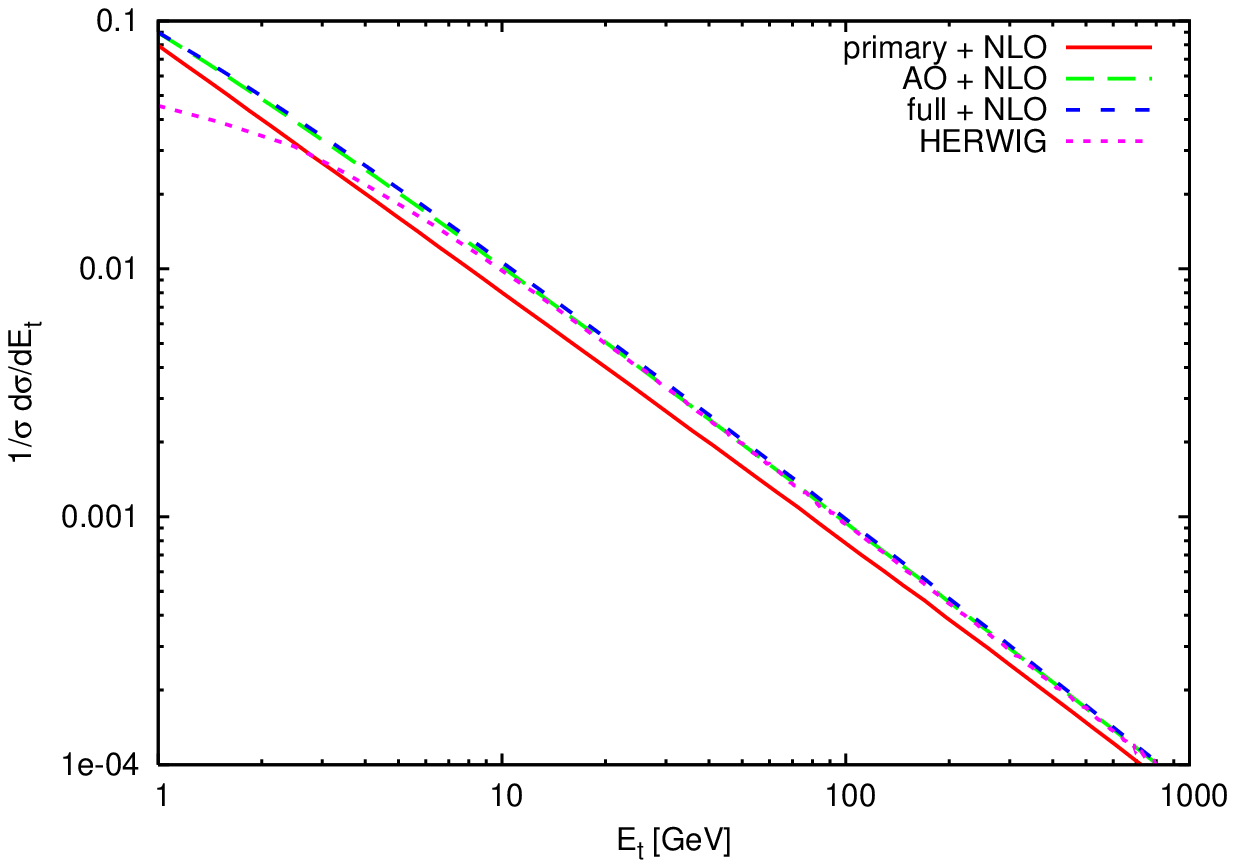,width=\textwidth}
    \caption{The distribution $\sigma^{-1} d\sigma/dE_t$ for a slice
      of $\Delta \eta =1$ and $Q=10^5~\GeV$ compared to parton shower
      results from HERWIG.}
\label{fig:Q1e5-herwig}
}

With the above observations in place, we start with the comparison to HERWIG 
which has a parton shower which
is ordered (in the soft limit) in angle and thus one would expect
results in line with those obtained via our AO model, introduced in
previous sections.  In Fig.~\ref{fig:Q1e5-herwig} we show the results
obtained from HERWIG compared to those from resummation for a rapidity
interval of unit width.  We note here that in order to obtain a
sensible behaviour for the resummed predictions at large $E_t$, it was
necessary to match the resummed results to exact fixed-order
estimates. We carried out the so-called log-$R$ matching~\cite{CTTW}
to both leading and next-to--leading order (obtained from the
numerical program EVENT2~\cite{event2}), but at the values of $E_t$ we
have shown here, no significant difference was observed. The curves
plotted in Fig.~\ref{fig:Q1e5-herwig} are matched to NLO while the
HERWIG results contain matrix-element corrections \cite{mecorr}.  We
observe that a very good agreement between HERWIG and the full and AO
curves is seen over a significant range of $E_t$ values.  We have also
included the value of the variable $t$ as a function of $E_t$ to
enable us to extrapolate our conclusions to lower centre-of-mass
energies.

\FIGURE{
  \epsfig{file=tval-Q1e5.eps,width=\textwidth}\\
  \epsfig{file=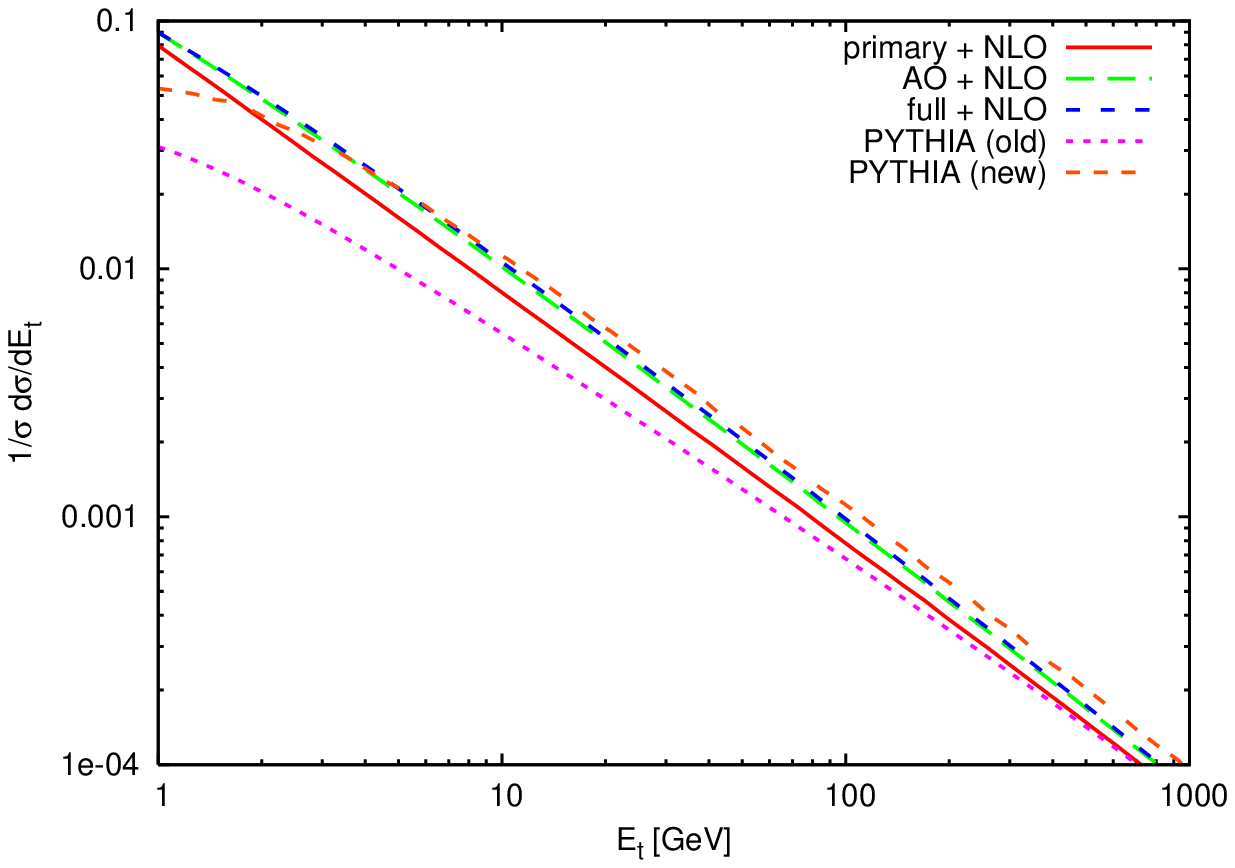,width=\textwidth}    
\caption{The distribution 
$\sigma^{-1} d\sigma/dE_t$ for a slice of 
$\Delta \eta =1$ and $Q=10^5~\GeV$ compared to parton shower results from 
PYTHIA.}
\label{fig:Q1e5-pythia}
}

The comparison to PYTHIA is shown in Fig.~\ref{fig:Q1e5-pythia}.  We
use version 6.3 and consider the old model, with showers ordered in
virtuality and forced angular ordering, as well as the new model,
where the emissions are ordered in transverse momentum.  We note that
the results obtained from PYTHIA with the new parton shower appear to
be in reasonable agreement with the resummed curves including
non-global logarithms, the situation being comparable to the quality
of agreement one obtains with HERWIG.  The same is not true for the
old PYTHIA shower and a significant disagreement between the result
there and the resummed curves is clearly visible.

In order to be more quantitative we focus on $E_t$ = 10 GeV which
corresponds to a value of $t=0.15$. Here we note that the difference
from the full resummed curve is respectively for HERWIG, PYTHIA (new)
and PYTHIA (old) approximately $-10 \%$, $+7.5 \%$ and $50 \%$. The
difference between a resummed primary contribution and the full
non-global result is, at the same value of $E_t$, $25 \%$. We would
then infer that if a variable of this type is chosen to tune for
instance PYTHIA with the old shower (with ordering in the invariant
mass) one includes potentially as much as $50 \%$ of the
leading-logarithmic perturbatively calculable contribution, to
model-dependent parameters and incalculable effects such as
hadronisation and the underlying event.

We have carried out our study for slices of different widths and
obtain comparisons with HERWIG that are generally satisfactory. The
same appears to be true of the new PYTHIA algorithm but here problems
seem to crop up as one increases the slice rapidity.  In
Fig.~\ref{fig:Q1e5py3} we present the comparison with both HERWIG and
PYTHIA, but for a slice width $\Delta\eta=3.0$.
\FIGURE{
    \epsfig{file=tval-Q1e5.eps,width=\textwidth}\\
    \epsfig{file=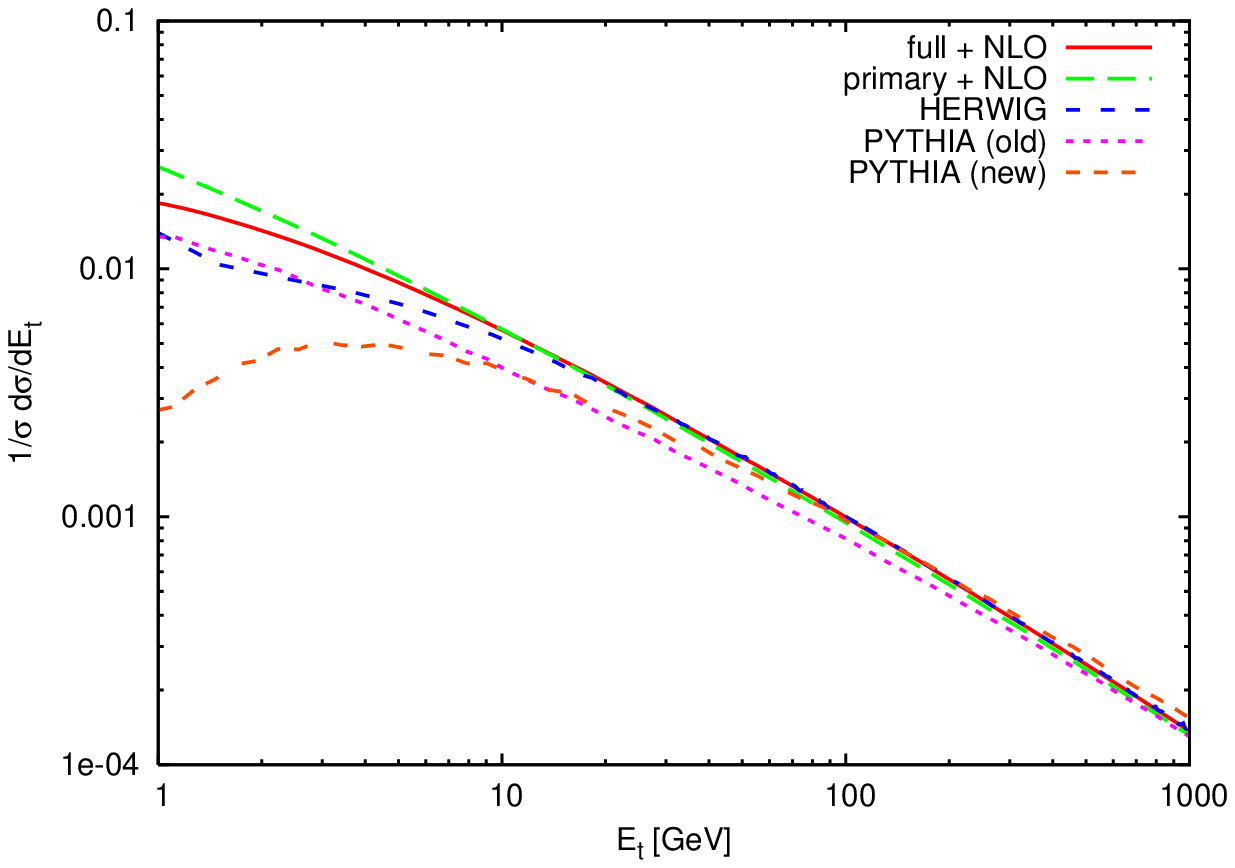,width=\textwidth}    
  \caption{The distribution $\sigma^{-1} d\sigma/dE_t$ for a slice of
    $\Delta \eta =3.0$ and $Q=10^5~\GeV$.}
\label{fig:Q1e5py3}
}
We observe that for a larger slice the new PYTHIA shower at lower
$E_t$ values yields a result that is significantly below all other
predictions.  The reason for this is not entirely obvious to us and we
would welcome further insight into this observation. We have also
carried out studies at other intermediate slice widths e.g. $\Delta
\eta =2.0$ and it appears that the new PYTHIA curve starts to deviate
from the resummed results at a value that is exponentially related to
the slice width. This may signal that the new ordering variable in
PYTHIA is perhaps not entirely satisfactory at large rapidities but as
we mentioned a more detailed study is required to draw firm
conclusions on this issue.

\section{Conclusions}
In this paper we have examined the role played by angular ordering in
the calculation of the leading single-logarithmic terms that arise for
non-global observables such as the away-from--jet energy flow.  While
it has been clear for some time \cite{DassalNG1,DassalNG2} that the
fully correct single logarithmic resummed result cannot be obtained
via use of angular ordering the question remained as to how much of
the full result for such an observable may be captured by using the
approximation of angular ordering. The reason this question arises in
the first place is mainly because angular ordered parton showers are
employed for example in Monte Carlo event generators such as HERWIG.
Given the importance of these event generators as physics tools it is
vital to understand the accuracy of the different ingredients thereof
(such as the parton shower).

While the accuracy of the parton showers is generally claimed to be
at least leading-logarithmic, this statement ought to apply only to those
observables where the leading logarithms are double logarithms, i.e.
both soft and collinear enhanced.
However, to the best of our knowledge, there has been no discussion yet
in the literature about non-global observables where the leading
logarithms may be single logarithms instead of double logarithms and
the accuracy of the parton showers in such instances. Since
observables of the type we discuss here (energy or particle flows in
limited regions of phase space) are often used in order to tune the
parameters of the Monte Carlo algorithms (see e.g.~\cite{CSH} for examples and
references), it is important to be at least aware of the fact the
perturbative description yielded by the parton shower, may in these
cases be significantly poorer than that obtained for instance for
global observables.  We have thus chosen one such observable and
carried out a detailed study both of the role of angular ordering as
well as the description provided by the most commonly used Monte Carlo
event generators HERWIG and PYTHIA, compared to the full
single-logarithmic result (in the large $N_c$ limit).

We find that in all the cases we studied, involving energy flow into
rapidity slices or patches in rapidity and azimuth, angular ordering
captures the bulk of the leading logarithmic contribution. This is a
comforting finding but there remains the issue of precisely how
angular ordering is embedded in the parton shower evolution for HERWIG
and PYTHIA.

For HERWIG where the evolution variable in the soft limit is the
emission angle one expects the agreement between parton shower and the
leading-log resummed descriptions to be reasonable and we find that
this is in fact the case.

In the case of the PYTHIA shower (prior to version 6.3) angular
ordering is implemented by rejecting non-angular-ordered
configurations in a shower ordered in virtuality. In this case it is
clear that the description of soft gluons at large angles will be
inadequate \cite{pytnew} and this feature emerges in our studies. From
this we note that a discrepancy of around 50\% could result while
comparing PYTHIA to the correct leading-log result. This difference
would be accounted for while tuning the parameters of PYTHIA to data
and must be borne in mind, for instance while making statements on the
tuning of the hadronisation corrections and the underlying event into
the PYTHIA model. This is because a tuning to energy flows would mean
that significant leading-logarithmic (perturbatively calculable)
physics is mixed with model-dependent non-perturbative effects which
does not allow for the best possible description of either.  Moreover,
the non-global effects are not universal and thus incorporating them
into the generic shower and non-perturbative parameters will lead to a
potentially spurious description of other (global) observables.  

The new PYTHIA shower, ordered in transverse momentum 
and with a more accurate treatment of angular ordering, does however give
a good description of the leading logarithmic perturbative physics,
comparable to that obtained from HERWIG. However, for large rapidity
slices we find that problems emerge in the description provided by
PYTHIA even with the new shower. The origin of these problems is not
entirely clear to us and we would welcome further insight here.
Hence, we strongly emphasise the need to compare the shower results
from HERWIG and PYTHIA while carrying out studies of observables that
involve energy flow into limited regions of phase space. Where this
difference is seen to be large, care must be taken about inferences
drawn from these studies about the role of non-perturbative effects,
such as hadronisation and the underlying event.  We believe that
further studies and discussions of the issues we have raised here are
important in the context of improving, or at the very least
understanding, the accuracy of some aspects of Monte Carlo based
physics studies.


\paragraph{Acknowledgements.}
We would like to thank Giuseppe Marchesini, Gavin Salam and Mike
Seymour for useful discussions. One of us (MD) gratefully acknowledges
the hospitality of the LPTHE, Paris Jussieu, for their generous
hospitality while this work was in completion.



\end{document}